\documentstyle[12pt]{article}
\def\la{{\langle}}
\def\ra{{\rangle}}
\def\wh{\widehat}

\newcommand{\beq}{\begin{equation}}
\newcommand{\eeq}{\end{equation}}
\newcommand{\beqa}{\begin{eqnarray}}
\newcommand{\eeqa}{\end{eqnarray}}

\begin{document}
\begin{center}
\Large{\bf Barrier traversal times using a phenomenological track
formation model$^*$
\vspace*{1.1cm}\\}
\large{J. P. Palao$^{a}$, J. G. Muga$^{a}$, S. Brouard$^{a}$ and
A. Jadczyk$^{b}$\vspace*{.1cm}\\}
{\it\normalsize  ${}^a$ Departamento de
F\'\i sica Fundamental y Experimental, Universidad de La Laguna,
Tenerife, Spain\\ 
${}^b$ Institute of Theoretical Physics, University of Wroclaw,
Poland\vspace*{2.cm}\\}

{\bf Abstract}
\end{center}
A phenomenological model for a measurement of ``barrier traversal
times'' for particles is proposed. Two idealized
detectors for passage and arrival   
provide entrance and exit times for the barrier traversal. 
The averaged traversal time is computed
over the ensemble of particles detected twice, before and after
the barrier. The ``Hartman effect'' can still be found when passage 
detectors that conserve the momentum distribution of the incident
packet are used.\vspace*{2cm}\\  
 $^*$Reference: Physics Letters A 233 (1997) 227-232\\
 (Electronic version with permission of Elsevier Science)

\newpage
\baselineskip 14pt
The temporal characterization of quantum mechanical tunnelling
traces back to early studies by McColl [\ref{MC}].
More recently a paper by B\"uttiker and Landauer [\ref{BL}] and
interest in the subject from various fields (as varied as nuclear
and molecular physics, cosmology or semiconductor physics) have
triggered a debate that has frequently dealt
with the very foundations of quantum mechanics [\ref{rev}]. 
The interpretation of the quantum mechanical formalism
and the wave-particle duality,
the quantization ambiguities, the relation between classical and
quantum mechanics, or the quantum ``measurement problem'' are some
of the ingredients of this research. These are all difficult and
not completely understood matters so, not surprisingly, answering
the question ``How long does it take to cross a barrier?'', i.e.,
defining a quantum traversal time has been
controversial. (We shall mainly discuss the general 
concept of ``traversal time'' instead of a more restrictive
``tunelling time''. However especial attention will be paid to 
tunnelling conditions in the calculations.)

In this problem the standard quantization
procedures are difficult to apply since only a limited  
number of classical trajectories
cross the selected region and ``continuous observation'' may be
required
for a measurement. Even so,  
many proposals exist that generalize in 
different formal or operational ways the classical concept of
traversal time to the quantum 
case. The debate on the barrier traversal time is essentially
a consequence of different conditionings and criteria, added
to the bare original question, that priviledge one quantum quantity
versus the others. As long as the conditioning is made explicit,  
to make clear that different versions of the 
original question are being answered, there is no fundamental
conflict among seemingly irreconcilable proposals. 
(Part of the theoretical work --using path integrals [\ref{Dim}] or 
a  projector approach [\ref{PRA94}]-- has been devoted
to develop comprehensive formalisms that allow to classify and relate 
many of the possible characteristic quantities.) However not all
aspects have yet been investigated. Only the totality of
conditionings or additional specificications 
exhausts the possible information about the barrier traversal 
in the temporal domain. Within this spirit we shall
investigate here a complementary aspect to those we have
previously examined [\ref{PRA94}-\ref{PRA96}], and to
experiments performed on electromagnetic waves
to measure ``Larmor times'' [\ref{BD}].      
The objective of this letter is to examine one {\it operational
definition of barrier traversal time} for particles.  
By ``operational'' we mean ``related to a specific 
experiment, possibly a ``Gedanken'' experiment.
%In fact when a quantization is problematic
%``experience'' has been traditionally invoqued to make a choice.
We shall model an idealized experimental setup 
inspired by an elementary ``classical'' receipe: In order to measure 
the transit time trough a spatial region the first entrance $t_1$ and
first exit times $t_2$ are measured and their difference
$\tau=t_2-t_1$ is evaluated {\it for each particle}.
In our case the spatial region includes a potential barrier 
and only the particles detected before {\it and} after the
barrier will be taken into account.  
If the experiment is repeated many times $\tau$ 
can be averaged and its statistial properties examined.
In a previous publication by Muga, Brouard and Sala [\ref{PLA92}]
a related approach was proposed for the quantum case:
An average entrance instant $\la t\ra^{\rm in}$ at $a$ and an average
exit instant $\la t\ra^{\rm out}$ at $b$ were defined in terms of
incident and outgoing current densities,
\beqa
\la t\ra^{\rm in}_a&=&\frac{\int_0^{t_c} J(a) t dt}{\int_0^\infty
J(a) t dt}\,,
\\
\la t\ra^{\rm out}_b&=&\frac{\int_0^\infty J(b) t dt}{\int_0^\infty
J(b) t dt}\,,
\eeqa
with a traversal time
$\tau_T\equiv\la t\ra^{\rm out}_a-\la t\ra^{\rm in}_b$
given by the difference between the two averages.
Here $J$ is the current density and it is assumed that $a$ is
far from the barrier so that the packet passes rightwards through
point $a$ before $t_c$, a time prior to the backwards reflected flow
after the collision. $\tau_T$ is in principle measurable but it
has a clear drawback since it is not the average of transit times for
{\it individual} particles. It is instead the difference between two 
averages of different nature. This is better understood in classical
terms:
The average entrance instant $\la t\ra^{\rm in}$ is operationally
defined for the ensemble of particles that arrive at the first detector
while the exit time is only defined
for a smaller set (those that arrive at the 
final detector). This definition in fact may lead to
negative values of $\tau_T$ in the classical and quantum cases
[\ref{LeOR},\ref{SSC}]. Classically the average entrance time may
be dominated by trajectories that are eventually reflected so that
$\la t\ra^{\rm in}$ can be very different from typical 
entrance times of the trajectories that eventually pass the barrier.
In this letter this inconsistency with the classical limit is
overcome by restricting the averaging to those particles that are
detected before {\it and} after the barrier. In general this approach  
implies a ``back reaction'' of the first detector that modifies the 
state. We accept this perturbation as a fact and investigate the
outcome of the described operational procedure, and the effect of
different detectors, in particular of those that minimize the back
reaction so that the momentum distribution of  
the initial packet is preserved. 
      
In general the particle$+$detector system  
involves many degrees of freedom 
and it is rarely modelled accurately. The objective of a phenomenological
model is to retain its essential aspects
with the aid of some adjustable set of parameters and in agreement with 
experiental facts. 
Our model does not specify the particular features of the detection at
a detailed experimental level but we have in mind {\it particle tracks}
similar to the ones produced in a bubble chamber or by means of
photographic plates. These tracks are characterized by a discrete 
set of macroscopic spots (two in our case)
originated at certain times (``clicks'')  considered as ``classical
events'' that result from the quantum particle passage or 
arrival. The particle is restricted to one dimensional spatial motion.
Specifically the effect of the detector associated with a given spot  
is simulated according to a track formation model proposed by 
A. Jadczyk and Ph. Blanchard using two basic elements [\ref{Arca}]:
An effective one-degree-of-freedom Hamiltonian and a  
modified projection postulate for the particle state 
after the first detection.    
\section{Model description}
The initial state of the particle is given by a wave function $\psi$  
associated with a preparation procedure. In operational terms, an
ensemble
of noninteracting particles, represented symbolically as $\{E_0\}$,
is sent towards the barrier -one particle at a time- from
the left with identical
specifications. (In our calculations the 
initial state at $t=0$ is a minimum-uncertainty-product Gaussian
centered at position $x=20$, momentum  $p=8$ and spatial variance
$9/4$, all quantities in atomic units. The potential barrier is a square 
barrier with ``height'' $V_0=50$ from $x=80$ to $x=80+d$ and the particle
has mass $m=1$.)

Two particle detectors $A$ and $B$
are located on both sides of a barrier potential,
at $x=a$ and $x=b$. The first one is a passage detector that
does not destroy the particle.
The second one is an arrival detector.
The translational degree of freedom of the particle, $x$,
is the only one represented explicitly.
A simplifying assumption is that only one of the two detectors is 
working at a time: When the particle is sent to the barrier
only $A$ is active. Detection of the particle at $A$ disconects
this detector and activates the second, $B$.  
\subsection{First detector: Probability of detection}
It can be 
proved using multichannel scattering theory techniques that the
incident channel amplitude (corresponding to translational motion of
the particle and the detector $A$ in its lower state) can
be represented by an effective Schr\"odinger equation with a
complex potential [\ref{Taylor}]. (In ``Event Enhanced Quantum Theory''
as described in [\ref{Arca}] the imaginary part of the potential is
deduced rigorously from the Lindblad form of the Liouville equation
that describes a coupling of the quantum system with a classical
detector.)
Here the effective Schr\"odinger equation is written as 
\beq
\label{S}
H\psi(x,t)=-\frac{\hbar^2}{2m}\frac{\partial^2}{\partial x^2}
\psi(x,t)+[V(x)+\Lambda(x)]\psi(x,t),
\eeq
where $V(x)$ represents the potential barrier
and the complex potential, $\Lambda$, is written as   
\beq
\Lambda(x)=-\frac{i}{2}g^2(x;a),
\eeq
with  
\beq
g(x,a)=s e^{-(x-a)^2/2\sigma^2}.
\eeq
The ``intensity'', $s$, and ``width'', $\sigma$,  of the detector
are adjustable parameters.

The norm of the incident channel, 
\beq
N(t)=\int_{-\infty}^{\infty} \psi^*(x,t)\psi(x,t)\,dx\,,
\eeq
decreases, due to the detector presence, from the initial value 
$N(0)=1$.
The total absorption $1-N(\infty)$ is the {\it efficiency}
of the detector.
It is not necessarily equal to one so the ensemble of  particles
detected at $A$,
$\{E_a\}$, is generally smaller than $\{E_0\}$. 
The normalized probability density for triggering
the detector at time $t_a$   
is proportional to the absorption rate $-dN/dt|_{t_a}$.
Normalizing with respect to the 
ensemble $\{E_a\}$ it is given by  
\beq\label{Pt1}
P(t_a|E_a)=\frac{dN(t_a)/dt_a}{\int_0^\infty dt\,dN(t)/dt}\,. 
\eeq
\subsection{Effect of detection on the wave functions} 
It will be assumed, within the spirit of a simplified 
phenomenological model, that after each detection (a ``click'') the
state of the particle can be effectively represented by a modified 
wavefunction. The true final states should be determined by a detailed
analysis of the interaction between the system and the detector.
Instead we shall later assume a physically motivated functional form. 
The ensemble of detected particles
can be represented by a statistical mixture of such states.
This is of course reminiscent of Von Neumman's projection postulate.  
However an important feature of a bubble chamber
track is that it does not look like a random walk. This cannot be
explained with a naive projection localizing the
particle position by means of position eigenstates,
since a position eigenstate has equal probability to expand in
any direction (erasing the memory of the state previous to the
measurement) so there would be no tendency to ionize atoms in the
direction of the dominant incident momentum [\ref{Ball}].
A modified projection postulate correcting this fact 
has been derived by Jadczyk and Blanchard. The wave function
resulting from a click at time $t_a$ and consistent with track formation
has a memory of the previous state and reflects also the detector 
properties. A simple expression satisfying these two conditions is 
[\ref{Arca}]  
\beq
\psi_{t_a}(x)=\frac{g(x)\psi(x,t_a)}{[\int_{-\infty}^\infty
g^2(x)|\psi(x,t_a)|^2\,dx]^{1/2}}\,,
\eeq
where $\psi(x,t_a)$ is the wave function evolved with the
Schr\"odinger equation (\ref{S}).
 
To determine the effect of the detector we have examined the
momentum average and its variance for the ensembles $\{E_0\}$ and
$\{E_a\}$. Averages over $\{E_a\}$ require some care since they imply a   
a double average: The first one (represented as $Q$) is a quantum
mechanical average using each wave packet $\psi_{t_a}$;
the second ($D$) is an average over the times of detection $t_a$
weighted by $P(t_a|E_a)$,  
\beq
\la p\ra_{E_a}=D Q p\equiv\int P(t_a|E_a)
\la\psi_{t_a}|\wh{p}|\psi_{t_a}\ra\, dt_a\,.
\eeq
Since there are two types of average different ``variances'' are
possible [\ref{ML},\ref{ML2}]. 
For the ensemble $\{E_a\}$ the important one is
$\Delta_{DQ}^2\equiv D Q [p^2-(D Q p)^2]$. (This is a
variance computed over detected particles regardless of their
detection time [\ref{ML}].) The average momentum is conserved well
(especially by weak detectors) except for very narrow detector widths.
For all detectors used in this work $ D Q p\approx \la p\ra_{E_0}$
better than 0.2$\%$.
However the ``momentum widths'' $\Delta_{DQ}$ (square root of variance) 
may change drastically with respect to the momentum width $\Delta_p$ of
the original packet. Fig. 1 shows that wider detectors tend to keep the
variance of the original state while narrow
detectors give very large variances. Weak detectors
(small $s$) conserve the variance
better than strong detectors (large $s$). In summary, in our model
weak and wide detectors are the best as far as conservation of
the momentum distribution of the original packet is concerned. 
They are however not very efficient, for $s=1$ the absorbed norm
goes from $0.05$ to $0.6$ in the $\sigma$-interval of Fig. 1. 
In comparison the full norm is absorbed for $s=10$.     
\subsection{The second (arrival) detector}
The second detector is assumed to be a perfect one as
described in [\ref{TAQM}], so that the full
transmitted packet is absorbed.
It is located at the right edge of the barrier. 
Let  $\{E_b\}$ be the ensemble of particles that produce two clicks
at times $t_a$ and $t_b$      
and $P(E_b|t_a)$ the transmittance of $\psi_{t_a}$, i.e., the 
fraction of the norm of $\psi_{t_a}$ that will be
transmitted and therefore detected at $B$ [\ref{trans}]. 
The probability for being detected at 
$B$ conditioned to having been detected at $A$ is 
\beq
P(E_b|E_a)=\int P(E_b|t_a)P(t_a|E_a)\, dt_a\,.
\eeq
Instead of using an expression similar to (\ref{Pt1})
the distribution of arrival
times $t_b$ for a perfect absorber can be approximated accurately 
by the (normalized) flux {\it without} absorber [\ref{TAQM}].
In particular, for a wave packet $\psi_{t_a}(x;t_a)$, the detection 
probability density at $t_b$ in $B$, conditioned to having been detected at 
$t_a$ in $A$ and restricted to the ensemble $\{E_b\}$, is given by  
\beq
P(t_b|E_b,t_a)=
\frac{J_{t_a}(b,t_b)}{\int J_{t_a}(b,t_b)dt_b}\,,
\eeq   
where $J_{t_a}$ is the flux for the state $\psi_{t_a}$. 
Using Bayes' rule the joint probability density for detection at $t_a$
in $A$ and $t_b$ at $B$ restricted to the ensemble $\{E_b\}$
is given by 
\beq
P(t_b,t_a|E_b)=
\frac{P(t_b|E_b,t_a)P(E_b|t_a)P(t_a|E_a)}
{\int P(E_b|t_a)P(t_a|E_a)dt_a}\,.
\eeq
Finally, the probability distribution of $\tau\equiv t_b-t_a$
is computed, for the ensemble $\{E_b\}$, by integrating 
over $t_b$ and $t_a$ with the delta function $\delta(t_b-t_a-\tau)$, 
\beq
P(\tau|E_b)=
\frac{\int P(t_a+\tau|E_b,t_a)P(E_b|t_a)P(t_a|E_a)dt_a}
{\int P(E_b|t_a)P(t_a|E_a)dt_a}\,.
\eeq
We have calculated average traversal times $\la\tau\ra_{E_b}\equiv
\int P(\tau|E_b) \tau d\tau$
versus the barrier 
width $d$ for two different weak detectors at $a$,
both with $s=1$. One of them, 
$A_{1}$, is a wide one and conserves well the momentum distribution of
$\{E_0\}$. The other one, $A_{2}$, is a narrow detector, and
produces a momentum variance which is approximately
ten times the initial one. The detector before the barrier 
is always put far from the barrier ($a= 50$) to compare with the 
type of Gedanken experiment performed in ref. [\ref{PLA92}],
so that the initial packet may pass through $a$ before
interacting significantly with the barrier, and $b$ is located at the
right barrier edge.  
Let $\tau_1$ and $\tau_2$ be the averages corresponding to 
using the two initial detectors $A_1$ and $A_2$.   
Figure 2 shows that the Hartman
effect, i.e. the fact that the average traversal time does not grow
with $d$ (actually it decreases slowly [\ref{SSC}]) can still be seen
with $A_{1}$ until a critical barrier width $d_c$ where the
``classical passage'' of momenta ``above'' the barrier starts to
dominate [\ref{PRA94},\ref{DM}]. When the narrow detector $A_{2}$ is
used the momentum variance is so large that the transmission is always
dominated by fast momenta well above the barrier (We have independently
checked this fact by calculating
the ratio between transmission due to energies above and below the
barrier energy.) so that the behaviour is the one expected classically,
i.e., a linear growth of $\tau_2$ with $d$.
Figure 2 also shows $\tau_T$, which is qualitatively very similar to 
$\tau_1$. The relation $\tau_T<\tau_1$ is due to the
two different ways the average is performed in the initial detector
and can be also understood on classical grounds.
The right front of the
incident packet is dominated by faster momenta and it contributes with
more
particles to the transmitted ensemble. For computing the later, no
distinction is made at $a$ between particles to be transmitted or not.
(The effect grows with $d$ until it saturates when the transmission is
purely above the barrier.)
Note that $\tau_T$ could be negative while the times defined in the
present work are always, by construction, strictly positive. 
The ``displacement'' of the curve $\tau_{\rm a}$ with respect to
$\tau_T$ (note the difference in the value of the critical barrier) 
is due to the slight difference in the momentum variances. 
           
In summary a two detector measurement of a particle traversal time
has been modelled.
Passage detectors conserving the initial wave packet momentum
distribution still show the Hartman effect.

Support by Gobierno Aut\'onomo de Canarias (Spain) (Grant PI2/95)
and by Ministerio de Educaci\'on y Ciencia (Spain) (PB 93-0578)
is acknowledged. J.P. Palao acknowledges an FPI fellowship from 
Ministerio de Educaci\'on y Ciencia.
\newpage
{\large \bf References}
\begin{enumerate}
\item\label{MC} L. A. McColl, Phys. Rev. 40 (1932) 621.
\item\label{BL} M. Buttiker and R. Landauer, Phys. Lett. 49 (1982) 1739
\item\label{rev}
E. H. Hauge and J. A. Stovneng, Rev. Mod.
Phys.  61 (1989) 917; 
M. B\"uttiker in  Electronic Properties of
Multilayers
and Low-Dimensional Semiconductor structures, ed. by 
Chamberlain J. M. et al (Plenum Press, New York, 1990) p. 297;
R. Landauer, Ber. Bunsenges. Phys.
Chem. 95 (1991) 404;
C. R. Leavens and G. C. Aers, in
Scanning Tunneling Microscopy and Related Techniques,
ed. by R. J. Behm, N. Garc\'\i a and H. Rohrer
(Kluwer, Dordrecht, 1990);
V. S. Olkhovsky and E. Recami, Phys. Rep. 214 (1992)
339; R. Landauer and T. Martin, Rev. Mod. Phys. 66 (1994) 217;
J. T. Cushing,  Quantum Mechanics (The
University of Chicago Press, Chicago, 1994).
\item\label{Dim} D. Sokolovski and J. N. L. Connor, 
Physical Review A 44 (1992) 1500.
\item\label{PRA94} S. Brouard, R. Sala and J. G. Muga,  Phys. Rev. A
49 (1994) 4312.
\item\label{PLA92} J. G. Muga, S. Brouard and R. Sala, 
Phys. Lett. A 167 (1992) 24.
\item\label{SSC} V. Delgado, S. Brouard and J. G. Muga, Solid State
Commun. 94 (1995) 979.
\item\label{TAQM} J. G. Muga, S. Brouard and D. Mac\'\i as,
Annals of Physics (NY) 240 (1995) 351.
\item\label{DM} V. Delgado and J. G. Muga, Annals of Physics
(NY) 248 (1996) 122.
\item\label{PRA96} S. Brouard and J. G. Muga,  Phys. Rev. A
54 (1996) 3055.
\item\label{BD} Ph. Balcou and L. Dutriaux, Phys. Rev. Lett. 78
(1997) 851.
\item\label{LeOR} C. R. Leavens, Solid State Commun. 85 (1993) 115;
89 (1993) 37.
\item\label{Arca}A. Jadczyk, Prog. Theor. Phys. 93 (1995) 631;
%particle tracks,events and quantum theory.
Ph. Blanchard and A. Jadczyk, Ann. d. Phys. 4 (1995) 583;
Ph. Blanchard and A. Jadczyk, Helv. Phys. Acta 69 (1996) 613.
\item\label{cal}The propagation of the wave packets is performed 
with the discretization algorithm described by S. E. Koonin 
in Computational Physics (Benjamin, Menlo Park, CA, 1985).
\item\label{Taylor} J. R. Taylor, 
Scattering  Theory (John Wiley, New York, 1972).
\item\label{Ball}L. E. Ballentine, Found. Phys.  11 (1990) 1329;
Phys. Rev. A 43 (1991) 9.
\item\label{ML} J. G. Muga and R. D. Levine, Mol. Phys. 67 (1989)
1209.
\item\label{ML2} J. G. Muga and R. D. Levine, Mol. Phys. 67 (1989)
1225.
\item\label{trans} J. G. Muga, S. Brouard and R. F. Snider,
Phys. Rev. A 46 (1992) 6075.

\end{enumerate}

\newpage
\centerline{\large Figure Captions}
{\bf Figure 1.} Square root of the momentum variance after detection,
$\Delta_{DQ}$, for $s=1$ (solid line) and $s=10$ (dashed line).
The dashed-dotted line is the reference value of the momentum variance 
for the original ensemble $\{E_0\}$.  

{\bf Figure 2.} Average traversal times versus barrier width $d$
evaluated
for (a) $s=1$, $\sigma=4.5$ (dashed line); (b) $s=1$, $\sigma=0.2$
(dashed-dotted line).
The average time $\tau_T$ is also represented (solid line).

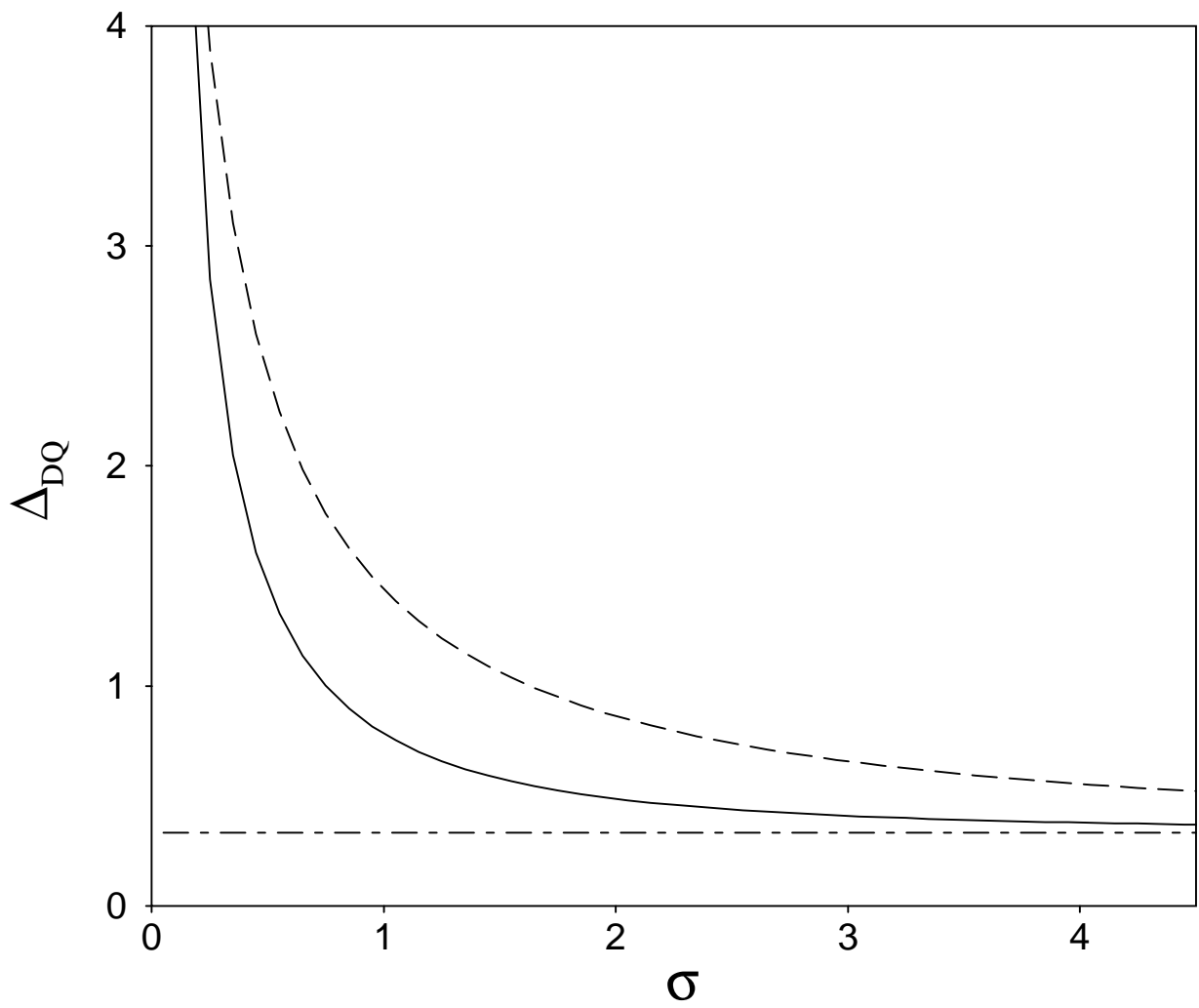
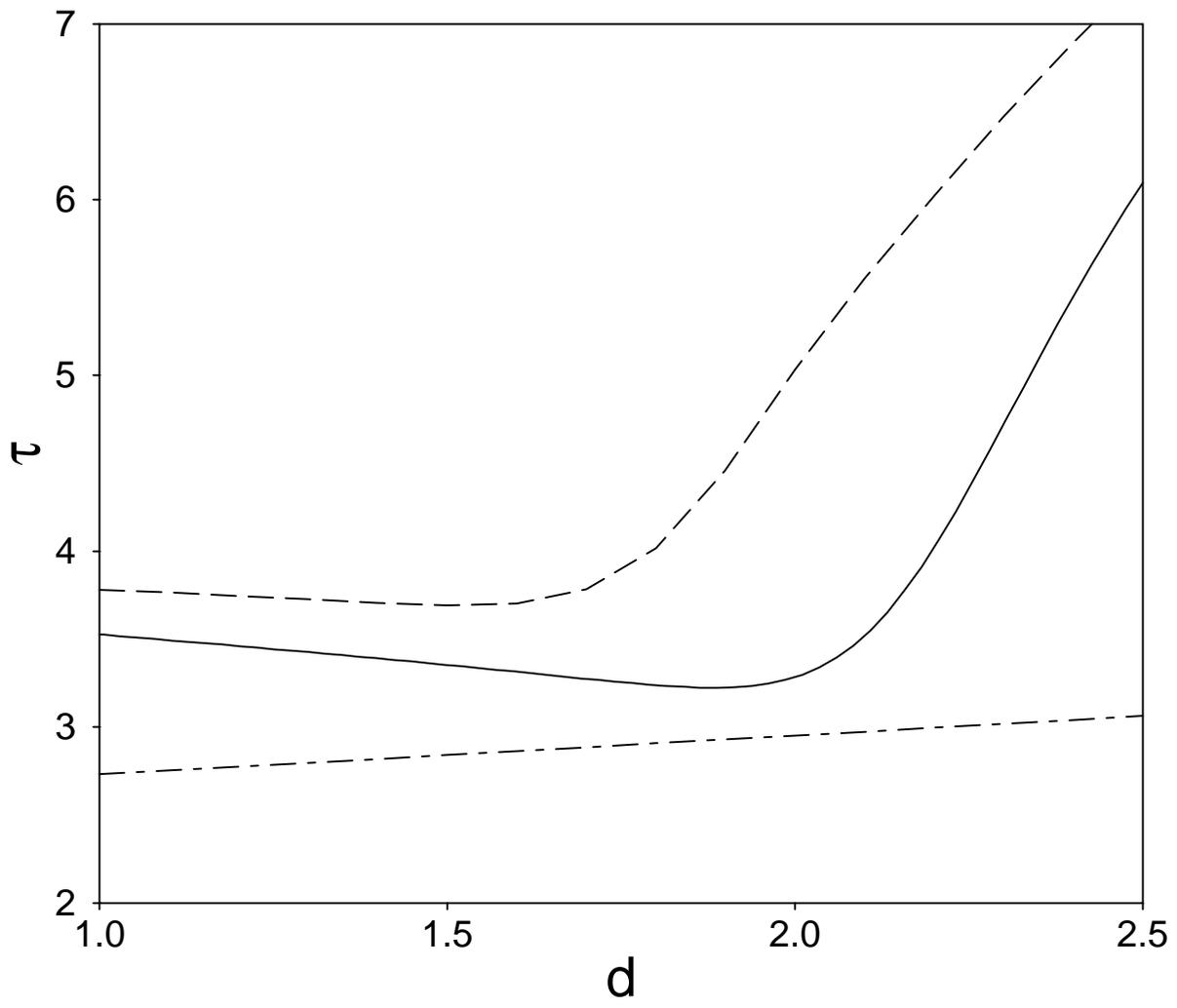
\end{document}